\documentclass[%
preprint,
amsmath,amssymb,
prb,
showpacs
]{revtex4-1}
\usepackage{graphicx}% Include figure files
\usepackage{dcolumn}% Align table columns on decimal point
\usepackage{bm}% bold math
\usepackage{braket}
\usepackage{color}
\usepackage{ulem} % strikethrough

\begin{document}
\title{Calculating branching ratio and spin-orbit coupling
    from first-principles: A formalism and its application to iridates}

\author{Jae-Hoon Sim$^1$}
\author{Hongkee Yoon$^1$} 
\author{Sang Hyeon Park$^1$}

\author{Myung Joon Han$^{1,2}$} \email{mj.han@kaist.ac.kr} %\email{mj.han@kaist.ac.kr}

\affiliation{$^1$Department of Physics, Korea Advanced Institute of Science and Technology (KAIST), Daejeon 305-701, Korea }
\affiliation{$^2$KAIST Institute for the NanoCentury, Korea Advanced Institute of Science and Technology, Daejeon 305-701, Korea }

\date{\today}

    %\affil[+]{these authors contributed equally to this work}
    %\keywords{Keyword1, Keyword2, Keyword3}

    %\usepackage{usebib} %related to citation
    %\bibinput{bibtex/DP_LS2}    %related to citation 
\pacs{75.70.Cn, 75.47.Lx, 71.15.Mb, 78.70.Dm}

\begin{abstract}
 We present a simple technique to calculate spin-orbit coupling,
 $\braket{ {\bf L}\cdot{\bf S}}$, and branching ratio measured in
 x-ray absorption spectroscopy.  Our method is for
 first-principles electronic structure calculation
 and its implementation is straightforward for any of standard
 formulations and codes.  We applied this technique to several
 different large spin-orbit coupling iridates.  The calculated
 $\braket{ {\bf L}\cdot{\bf S}}$ and branching ratio of a prototype $j_{\rm
     eff}$=1/2 Mott insulator, Sr$_2$IrO$_4$, are in good agreement with
    recent experimental data over the wide range of
    Rh-doping.  Three different double perovskite
    iridates (namely, Sr$_2$MgIrO$_6$, Sr$_2$ScIrO$_6$, and
    Sr$_2$TiIrO$_6$) are also well described.
    This technique can
    serve as a promising tool for studying large spin-orbit coupling
    materials from first-principles and for understanding experiments.
\end{abstract}

%75 Magnetic properties and materials
%%71.70.Ch	Crystal and ligand fields

%73. Electronic structure and electrical properties of surfaces, interfaces, thin films, and low-dimensional structures
%%73.20.-r	Electron states at surfaces and interfaces

%71. Electronic structure of bulk materials
%%71.15.Mb	Density functional theory, local density approximation, gradient and other corrections

%78. Optical properties, condensed-matter spectroscopy and other interactions of radiation and particles with condensed matter
%%78.70.Dm	X-ray absorption spectra
\flushbottom
\maketitle
\thispagestyle{empty}

%\noindent Please note: Abbreviations should be introduced at the first mention in the main text – no abbreviations lists. Suggested structure of main text (not enforced) is provided below.

\section{Introduction}

Recently the role of spin-orbit coupling (SOC) in solids has attracted
tremendous attention.  In many cases, SOC drastically changes the
electronic band structure and results in a fundamentally different
material property. A class of materials, called topological
insulators, is an outstanding example \cite{hasan_textitcolloquium_2010,qi_topological_2011}.  SOC
can also play together with on-site electronic correlation, $U$, as
often found in $5d$ transition-metal oxides.  In iridates, for
example, the cooperation of SOC and $U$ drives materials to be a novel
`$j_{\rm eff}$=1/2 Mott insulator' \cite{kim_novel_2008,kim_phase-sensitive_2009}.  
Due to the characteristic hopping integrals caused by $j_{\rm eff}$=1/2 nature
(instead of $S$=1/2), some interesting new possibilities have been
proposed and still under active investigations \cite{shitade_quantum_2009,pesin_mott_2010,jackeli_mott_2009,wang_twisted_2011}.
The basically same features can also be found in the non-oxide $4d$ and
$5d$ transition-metal compounds \cite{kim_spin-orbital_2014}.

The spin-orbit Hamiltonian is represented by
$\lambda\braket{ {\bf L}\cdot{\bf S}}$.  While $\lambda$ is known from the
atomic nature of a given species, the direct estimation of 
$\lambda\braket{ {\bf L}\cdot{\bf S}}$ is not always straightforward from experiment nor
by theoretical calculation. For topological insulators, the observed
band structure ({\it e.g.}, by angle-resolved photoemission
spectroscopy (ARPES)) is regarded as a strong evidence of the
characteristic band dispersion caused by SOC \cite{hasan_textitcolloquium_2010,qi_topological_2011}. For
iridates, the data from resonant x-ray magnetic scattering (RXMS) and/or
resonant inelastic x-ray scattering (RIXS) have been accepted as a
confirmation of the novel SOC physics because the interpretation of
the data seems consistent only with theoretical models that take
strong SOC into account \cite{kim_phase-sensitive_2009,kim_magnetic_2012}. However it
is noted that sometimes a different interpretation can be made and
then the conclusion might be changed
(for an example of iridates, see Ref.~\cite{moretti_sala_resonant_2014}).
Further, from the theoretical point of view, it is unsatisfactory 
that there is no simple and well-defined way to directly calculate SOC strength 
and to compare with experiments.  
In the standard first-principles calculations,
$\lambda$ can be calculated when the atomic wavefunctions are
constructed by solving the relativistic Dirac
equation. $\braket{ {\bf L}\cdot{\bf S}}$, however, is not just determined
by atomic nature but depends on the electronic structure of solids.

\begin{figure}
    \centering
    \includegraphics[width=0.6\linewidth]{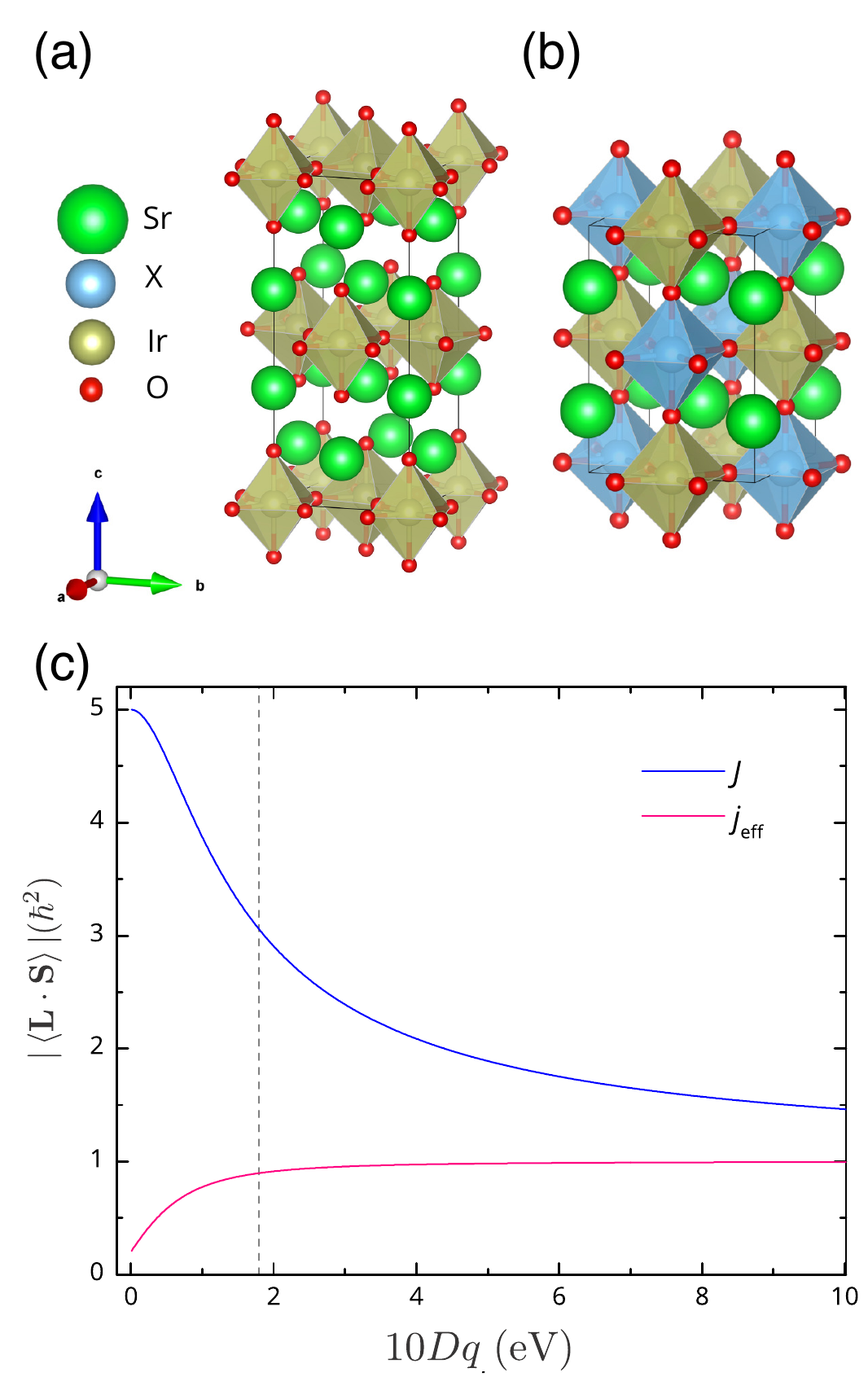}
    \caption{The unitcell structure of (a) Sr$_2$IrO$_4$ and (b) double perovskite. The green, light blue, light brown, and red spheres represent Sr, $X$, Ir and O, respectively ($X$ = Mg, Sc, Ti). The $X$O$_6$ and IrO$_6$ cage is shaded in blue and brown, respectively. (c) The expectation value of spin-orbit coupling calculated in the $d^5$ atomic limit as a function of $10Dq$. 
    The blue ($\braket{{\bf L}\cdot {\bf S}}_J$) and
    red ($\braket{{\bf L} \cdot {\bf S}}_{j_{\text{eff}}}$) line corresponds to the result obtained 
    by using both $t_{2g}$ and $e_g$ orbitals and only $t_{2g}$ orbitals, respectively.
    }
    
    \label{fig:Fig1}
\end{figure}

In this paper, we point out that the calculation of
$\braket{ {\bf L}\cdot{\bf S}}$ can be performed in a simple and straightforward way
within the standard first-principles framework and be directly compared with
experiment.  One possible reason that the calculation of
$\braket{ {\bf L}\cdot{\bf S}}$ has not been often made from first-principles
may be partly because of no direct reference data available from the
experimental side.  We note that the branching ratio, typically
measured in x-ray absorption spectroscopy (XAS), can be used to
estimate the strength of SOC.  Instead of calculating XAS spectrum itself, a
simple technique can be used to directly calculate branching ratio through 
$\braket{ {\bf L}\cdot{\bf S}}$.  Our formalism is
implemented into our localized pseudo-atomic orbital (PAO) basis code and
applied to several different iridium oxide compounds. For a
$j_{\rm eff}$=1/2 system, Sr$_2$IrO$_4$ (see Fig.~\ref{fig:Fig1}(a)),
we considered Rh doping
(namely, Sr$_2$Rh$_{x}$Ir$_{1-x}$O$_4$) and found that the calculated SOC and
branching ratio are in good agreement with
XAS data over the wide range of doping ratio $x$.
The iridate
double perovskites (see Fig.~\ref{fig:Fig1}(b)),
Sr$_2$$X$IrO$_6$ ($X$: Mg, Sc, Ti), are also calculated and the results are
in good agreement with recent experiments.  We emphasize the formalism
and implementation are simple enough to be adoptable for any type of
first-principles code and method. This technique can be a useful tool
for study large SOC materials by providing a direct estimation of
$\braket{ {\bf L}\cdot{\bf S}}$ and branching ratio.

\section{COMPUTATION METHOD}

\subsection{Calculation details}
For the electronic structure calculations, we used 
OpenMX software package \cite{_openmx,ozaki_numerical_2004,ozaki_variationally_2003}
which is based on the linear combination of numerical PAO
and norm-conserving pseudopotential \cite{ozaki_variationally_2003}.
The cutoff radii for Sr, Ir, Rh, Mg, Sc, Ti, and O are
10.0, 7.0, 7.0, 7.0, 7.0, 7.0, and 5.0 a.u. respectively.
The Perdew-Burke-Ernzerhof (PBE) exchange-correlation functional 
\cite{perdew_generalized_1996} and 300 Ry energy cutoff.
$5 \times 5 \times 2$ and $9 \times 9 \times 7$ k-meshes were taken for Rh-doped Sr$_2$IrO$_4$ and double perovskites, respectively. 
have been used. 
The SOC was treated within a fully relativistic
$j$-dependent pseudopotential scheme in the non-collinear methodology
\cite{_openmx}.
The on-site electronic correlations were taken into account within
DFT+$U$ formalism \cite{dudarev_electron-energy-loss_1998,han_$mathrmon$_2006}.
The reasonable value of $U$ may be about  2.0 -- 3.0 $eV$
as noticed by the previous studies on Sr$_2$IrO$_4$ and Ba$_2$IrO$_4$
\cite{kim_novel_2008,zhang_effective_2013,martins_reduced_2011,arita_textitab_2012}.
Throughout the manuscript, we present $U_\text{eff} \equiv U-J = 2.0 $ eV 
results as our main data both for Rh-doped Sr$_2$IrO$_4$ and double
perovskites. After scanning the region of $U_\text{eff} =$ 2.0 -- 3.0 eV,
we found that any of our conclusion does not change by choosing different $U$ values.
For Rh-doped iridates, the lattice constant and internal coordinates
are optimized with the force criteria of 0.01 eV/{\AA}.
For double perovskites, we used the experimental lattice parameters \cite{kayser_crystal_2014,jung_high_1995}
of $a = 3.958 ${\AA} (Sr$_2$MgIrO$_6$), $4.007${\AA} (Sr$_2$ScIrO$_6$)
and $3.927${\AA} (Sr$_2$TiIrO$_6$).

\subsection{Formalism} 

In this section, we present our formalism to calculate
$\braket{ {\bf L}\cdot{\bf S}}$ and branching ratio from
first-principles.
The localized atomic orbitals are assumed to be the basis set in the below. However,
it is straightforward to extend our method to any other type.
We used our pseudopotential-based DFT (density functional theory) package,
OpenMX \cite{_openmx},
which takes the linear combination of numerical PAO basis
\cite{ozaki_variationally_2003,ozaki_numerical_2004}.
The single particle energy eigenstate is decomposed into PAO;
	$\ket{\psi_{n{\bf k}}}=\sum_{i,\alpha} c_{\alpha,i}^{n,{\bf k}}
	\ket{\phi_{\alpha,i}}$
 where $\ket{\psi_{n{\bf k}}}$ is Khon-Sham
eigenstate with momentum ${\bf k}$, $n$ the band index, and 
$\ket{\phi_{\alpha, i}}$ is PAO with orbital index $\alpha$ at
position $R_i$.
With $J=|{\bf L} + {\bf S}|=5/2$, $ 3/2$ state as a basis set
for a given Ir-$5d$ orbitals,
\begin{equation}
%\begin{split}
\ket{\psi_{n {\bf k}}}
=   \sum_{m_J=-5/2}^{5/2}a^{n{\bf k}}_{m_J}    \ket{J=5/2, m_J}_\textrm{Ir}  
 + \sum_{m_J=-3/2}^{3/2}b^{n{\bf k}}_{m_J}   \ket{J=3/2, m_J}_\textrm{Ir} 
 + \sum_{(i,\alpha) \neq (\textrm{Ir}, d)} c^{n{\bf k}}_{\alpha,i}  \ket{\phi_{\alpha,i}}.
\label{eq:basis}
%\end{split} 
\end{equation}
Now the expectation value of ${\bf L}\cdot{\bf S}$ is estimated 
within this basis as follow:
\begin{equation}
\begin{split}
\braket{{\bf L}\cdot {\bf S}} =& 
\sum_{n{\bf k}}^{\text{occ}}  \bra{\psi_{n\bf k}} {\bf L}\cdot {\bf S} \ket{\psi_{n\bf k}}\\ 
=& \sum_{\epsilon_{n{\bf k}}< \epsilon_F}  \sum_{m_J}
(1.0 \times |a^{n{\bf k}}_{m_J}|^2 -1.5 \times |b^{n{\bf k}}_{m_J}|^2),
\end{split}
\label{eq:LS_DFT}
\end{equation}
where $1.0$ and $1.5$ are the eigenvalue of the ${\bf L \cdot S}$ operator
for $J=5/2$ and $3/2$, respectively.

Note that it is crucial to use the $J$ state as a basis and the
$j_{\text{eff}}$ is not suitable
although it is often adapted to describe the low
energy electronic structure of iridates. 
Fig.~\ref{fig:Fig1}(c) shows the calculated $\braket{{\bf L}\cdot {\bf S}}$
as a function of crystal field splitting $10Dq$ 
with the basis set of total angular momentum $J$ eigenstates (blue)
and $j_{\text{eff}}$ states (red). For a reasonable value of $10Dq \approx 1.8$eV 
for the iridates \cite{haskel_pressure_2012},
the two lines differ significantly due to the coupling between $t_{2g}$ and
$e_g$ states. It is noted that, even in a large $10Dq$ limit,
$\braket{{\bf L}\cdot {\bf S}}_J $ and $\braket{{\bf L}\cdot{\bf S}}_{j_{\text{eff}}}$
can noticeably differ from each other.
Here $\braket{{\bf L}\cdot {\bf S}}_J $ is the same quantity with $\braket{{\bf L}\cdot {\bf S}}$ in Eq.(\ref{eq:LS_DFT}), and
	$\braket{{\bf L}\cdot {\bf S}}_{j_{\text{eff}}}$ is obtained from only $t_{2g}$ space taken into account.

The `line strength' $L_j$ can be expressed as the expectation value of an operator
\begin{equation}
\hat P_j \equiv \sum_{\lambda,q} \hat D_q \ket{\lambda j}\bra{\lambda j} \hat D_q,
\end{equation}
where $D_q$ is the dipole operator with polarization $q$, $j$ is the total angular momentum of the core hole, and $\lambda$ denotes all quantum numbers other than $j$.
The sum is taken over the $q=-1,0,1$ for the isotropic `line strength' considered.

Branching ratio can also be estimated without calculating the full XAS spectra.
We first note that the relative intensity of $L_3$ and $L_2$ edge `white lines'
can be related to the spin-orbit coupling via
\begin{equation}
\frac{L_j}{L_3+L_2} = \frac{2j+1}{2(2l_c+1)} \pm A(l_c,l_v,n_h) \braket{{\bf L} \cdot {\bf S}},
\label{eq:BR_LS}
\end{equation}
where $l_v$, $l_c$, and $j=l_c\pm1/2$ refers to the orbital angular momentum of valence electron,
the orbital angular momentum of core hole, 
and the total angular momentum of the core hole,
respectively
\cite{van_der_laan_local_1988,thole_linear_1988}.
The number of holes in valence orbitals is denoted by $n_h$.
In our case,
\begin{equation}
A(l_c,l_v,n_h)=\frac{2-l_v(l_v+1)-l_c(l_c+1)}{l_v(l_v+1)(2l_c+1)(n_h)}=-\frac{1}{3n_h}.
\end{equation}
This relation is expected to be exact for the dipole transition in which the core-hole interaction
with the valence electrons is small enough in comparison to 
the spin-orbit interaction of the core hole. Therefore our case of iridates
is suitable for this formalism to be applied.
From Eq.(\ref{eq:BR_LS}), the branching ratio can be written as
\begin{equation}
I_{L_3} / I_{L_2} =
\frac{2n_h -\braket{{\bf L} \cdot {\bf S}}}{n_h+\braket{{\bf L} \cdot {\bf S}}}
= \frac{2-r}{1+r}
\end{equation}
with $r= \braket{{\bf L} \cdot {\bf S}} / n_h$.
Therefore the branching ratio can be estimated by calculating the number of holes (which is straightforward in the electronic structure calculation) and $\braket{{\bf L} \cdot {\bf S}}$ (which can also be estimated as described above).

Due to recent progress, calculating XAS spectra from first-principles
becomes feasible.
Some codes are already available for this capability
\cite{giannozzi_quantum_2009,pardini_first-principles_2012}. However the
calculation of the whole spectra is quite demanding in general. For example, a
certain type of pseudopotential (such as projector augmented wave)
should be prepared for describing the core holes. Also, the
generalization to the non-collinear spin configuration space is
sometimes not well prepared while the
non-collinear spin order is actually stabilized in many of large SOC materials 
as in the case of Sr$_2$IrO$_4$. Our technique has a clear advantage in these
regards. First of all, it is much simpler in the implementation and calculation,
and does not require any special type of pseudopotential. One can
just use the original code as it is and the only required information
is the final band structure that is properly transformed into
$J$-space. In spite of its simplicity, the quantitative comparison can
still be made with experiment through the branching ratio, and the
direct estimation of SOC is also provided although the full XAS spectra
is not accessible.

\section{Result and Discussion: Application to Iridates}
\subsection{Rhodium-doped Sr$_2$IrO$_4$}

\begin{figure}
    \centering
    \includegraphics[width=0.99\linewidth]{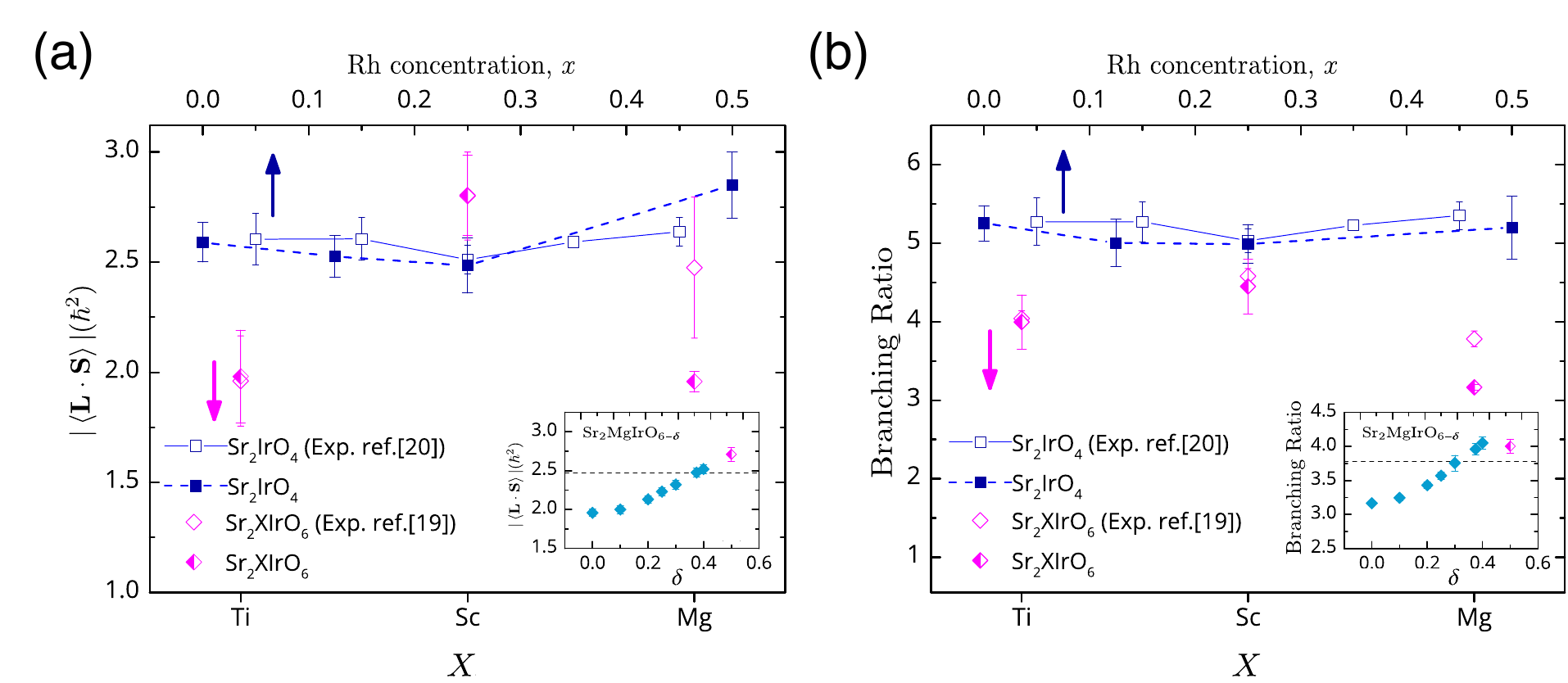}
    \caption{            
        {The calculated values (filled or half-filled symbols) 
            of (a) $\braket{{\bf L}\cdot{\bf S}}$ and (b) branching ratio in comparison with experiments (open symbols).
            The upper (corresponding to the dark blue symbols) and lower axis (magenta symbols) refers to 
            Sr$_2$Rh$_{x}$Ir$_{1-x}$O$_4$ and Sr$_2$$X$IrO$_6$
            ($X$ = Mg, Sc, Ti), respectively. The error bars
            for experimental data are
            taken from the original papers. The error bars in the calculation results
            reflect the ambiguity related to the numerical parameters and other computation details
            (see the main text for more details). Inset:
            The  effect of oxygen vacancy has been simulated for Sr$_2$MgIrO$_{6-\delta}$. The horizontal dashed lines refer to the experimental values. The filled (blue) symbols  represent the result of rigid band shift and the half-filled (magenta) symbols are the result of the supercell calculation with oxygen removal }
    }
    \label{fig:LS_BR}
\end{figure}

The calculation results of Rh-doped iridates,
Sr$_2$Rh$_{x}$Ir$_{1-x}$O$_4$, are summarized in Fig.~\ref{fig:LS_BR}
(see the filled blue squares; corresponding to the upper $x$-axis) 
where XAS data is also presented (open blue squares).The good agreement between calculation
and experiment is clearly noticed over the wide range of Rh-doping ratio, $x$. 
Note that the error can be caused in both theoretical
and experimental estimation as marked by the error bars. In
calculations, the one important source of error is the range of
integration; namely, how to deal with the small portion of
Ir-$5d$ states hybridized with oxygen states while the major Ir
peaks are clearly identified. 
%In our calculation, we used the same
% integration range for for all $x$;
% $-$4.5 -- $-$2eV and 5.5 -- 10eV.
According to our estimation, this intrinsic ambiguity can cause
the deviation of branching ratio by up to $\pm$0.25 which is
$\lesssim 5$\%. 
Counting the number of holes (or electrons) in Ir-5d orbitals is
another source of errors. This is related to the long-standing
issue of charge decomposition in the electronic structure calculation,
and the number of holes
depends on the electron counting method.
In this study, we used the standard Mulliken charge analysis.
For experimental data, we simply take the error limits presented
in Ref.~\cite{laguna-marco_electronic_2015,chikara_sr_2ir_1_2015}.

Note that the $\braket{ {\bf L}\cdot{\bf S}}$  and branching ratio
 are basically unchanged over 
the wide range of $x$, and clearly larger than 
the Rh value of $\sim$0.8 and $\sim$3, respectively \cite{chikara_sr_2ir_1_2015}.
This result is therefore in
contrast to the previously suggested picture of `SOC
tuning' in which Rh-doping is assumed to reduce the SOC strength of Ir sites
\cite{lee_insulator-metal_2012,qi_spin-orbit_2012}.
It is one example to show the importance of calculating SOC  
from the realistic electronic structure.

\subsection{Iridium oxide double perovskites}

Another system we take to test our method is iridate double perovskites, Sr$_2X$IrO$_6$
($X$: Mg, Sc, Ti).
This series of materials are
studied recently with XAS \cite{chikara_sr_2ir_1_2015} while no
theoretical investigation has been reported yet.
Among many different double perovskite iridates, we chose 
$X$= Mg, Sc, and Ti in which
$X$ has $d^{0}$ configuration, and therefore we could avoid the additional ambiguity in 
determining $U$ values for $X$ sites.
The nominal Ir valence in
these compounds are $6+$, $5+$, and $4+$ for Sr$_2$MgIrO$_6$,
Sr$_2$ScIrO$_6$, Sr$_2$TiIrO$_6$, respectively, 
serving as a good test case to check the reliability of our method.

\begin{figure}
    \centering
    \includegraphics[width=0.6\linewidth]{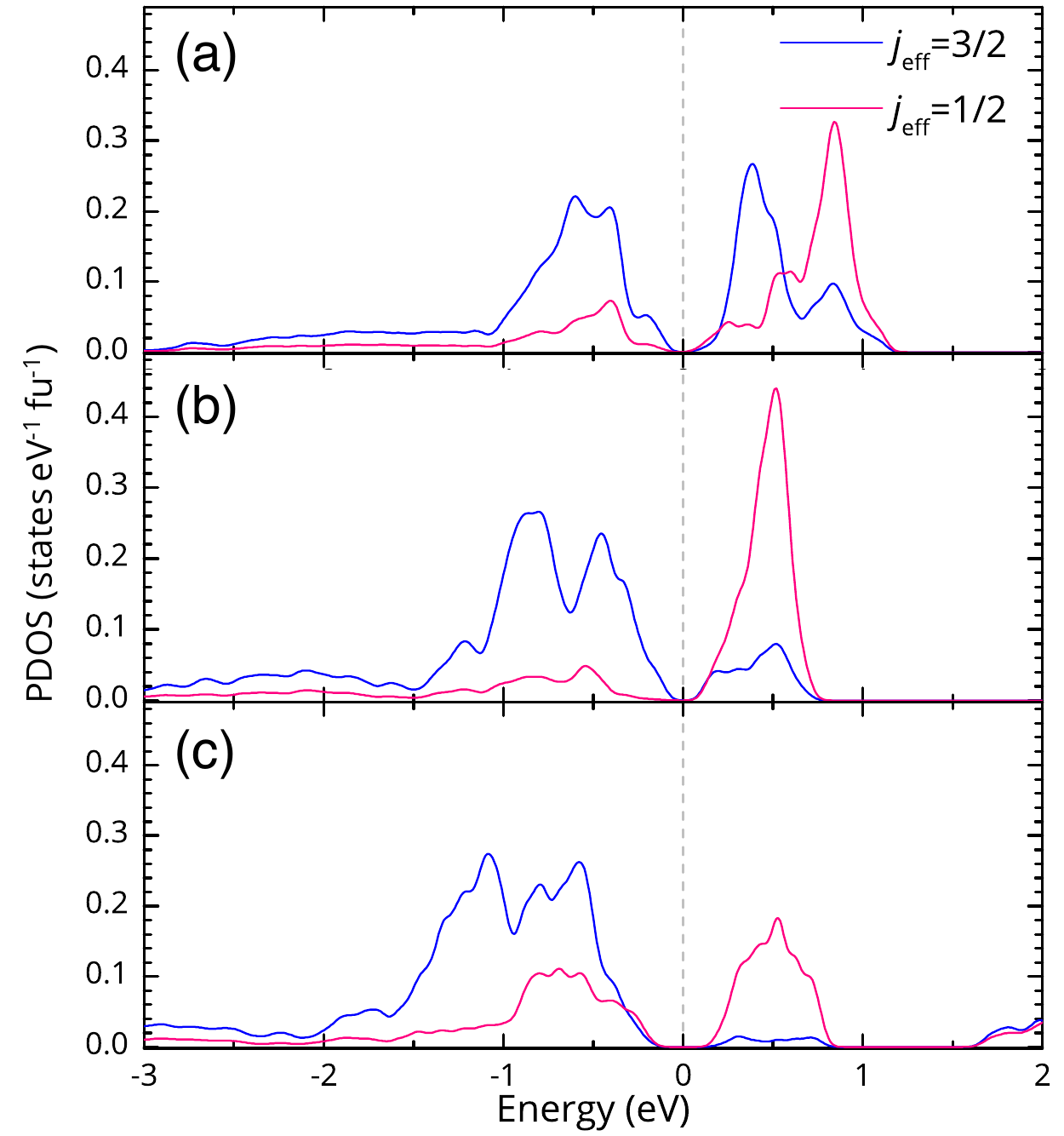}
    \caption{The calculated PDOS for (a) Sr$_{2}$MgIrO$_{6}$, (b) Sr$_{2}$ScIrO$_{6}$ and (c) Sr$_{2}$TiIrO$_{6}$, corresponding to Ir valence of 6+, 5+, and 4+, respectively. The blue and red lines represent $j_{\rm eff}$=3/2 and 1/2 states, respectively.}
    
    \label{fig:DOS}
\end{figure}

The calculation results of projected density of states (PDOS) is presented in Fig.~\ref{fig:DOS}. 
The so-called effective total angular momentum $j_\text{eff}$-character is well
identified ($j_{\rm eff}$=1/2 and 3/2 is
in red and blue color, respectively).
The gradual increase of Ir-$5d$ band filling is clearly observed as 
we go from Mg ($\frac{3}{4}$-filled $j_{\rm eff}$=3/2) to Sc (fully-filled $j_{\rm eff}$=3/2)
and to Ti (half-filled $j_{\rm eff}$=1/2).
The Ti case has the same nominal valence with 
Sr$_2$IrO$_4$. 
Among these three compounds, Sr$_2$MgIrO$_6$ is reported first
by Jung and Demazeau \cite{jung_high_1995}
and its semiconducting behavior in the transport \cite{jung_high_1995}
and antiferromagnetic ordering below $T_N=74$K  \cite{kayser_crystal_2014}
are consistent with our results.

The results of $\braket{ {\bf L}\cdot{\bf S}}$ and branching ratio
are summarized in Fig.~\ref{fig:LS_BR}(a) and (b), respectively
(see the magenta diamond symbols corresponding to the lower axis).
Again, the overall good agreement with XAS data is clearly noticed.
Here we also note several possible sources of deviation, when compared to
the experiments, such as the material dependent $U$ values, long PDOS
tails due to the hybridization, and the electron number
counting with the Mulliken analysis. 
Considering the uncertainties related to all of these factors,
the agreement within the error bar is quite impressive.
According to the previous studies \cite{jung_high_1995,kayser_crystal_2014},
it is noted that the amount of oxygen vacancy is likely non-negligible  especially
for $X$=Mg sample (Sr$_2$MgIrO$_{6-\delta}$: $\delta \sim$ 0.35 at 1 bar). It is another factor 
that can cause the difference between
experiment and calculation.
In fact, the difference 
is distinguishable (considering the  
error bars) only in the $X$=Mg case. In order to see the effect of oxygen vacancy,
we performed the rigid-band shift ($\delta$= 0, 0.1, 0.2, 0.3, 0.4) and the supercell calculation ($\delta$=0.4). In
the former, it is assumed that the role of vacancy is just electron doping.
The results are summarized as the insets of
 Fig. \ref{eq:BR_LS}. An excellent agreement with experiments
 for both $\braket{ {\bf L}\cdot{\bf S}}$ and branching ratio
 is clearly noticed.

While our main results are obtained from 20-atom unitcells with antiferromagnetic order,
we also performed
the 10-atom cell calculations which correspond to ferromagnetic 
order. The calculated SOC and branching ratio are not
noticeably different. It indicates that the effect of magnetic order
and structural distortion 
is not significant. It is also found that 
the $U$ dependence is not significant in the range we considered.

\section{Summary} 

We introduce a technique to calculate SOC 
and branching ratio. 
It is simple enough to be adoptable basically for any type of first-principles calculation
methods and formalisms. 
Large SOC iridates were taken to be a test case for
this method. The calculation results 
of Rh-doped  Sr$_2$IrO$_4$ and double perovskites,
Sr$_2$$X$IrO$_6$, are shown to be
in good agreement with recent XAS experiments.
This technique can serve as a new promising tool for studying large
SOC materials from first-principles and for
understanding related experiments.

\section{Acknowledgements} %(not compulsory)}

%Acknowledgements should be brief, and should not include thanks to anonymous referees and editors, or effusive comments. Grant or contribution numbers may be acknowledged.
MJH is greatly thankful to Michel van Veenendaal for helpful
discussion.  This work was supported by Basic Science Research Program
through the National Research Foundation of Korea (NRF) funded by the
Ministry of Education (2014R1A1A2057202).  The computing resource is
supported by National Institute of Supercomputing and Networking /
Korea Institute of Science and Technology Information with
supercomputing resources including technical support
(KSC-2014-C2-015).

\end{document}